\documentclass[twocolumn]{aastex63}

\usepackage{amsmath}
\usepackage{amssymb}
\usepackage{graphicx}
\usepackage{epsfig}
\usepackage{color}
\usepackage{natbib}

\def\gtaprx {\lower .1ex\hbox{\rlap{\raise .6ex\hbox{\hskip .3ex
	{\ifmmode{\scriptscriptstyle >}\else
		{$\scriptscriptstyle >$}\fi}}}
	\kern -.4ex{\ifmmode{\scriptscriptstyle \sim}\else
		{$\scriptscriptstyle\sim$}\fi}}}
\def\ltaprx {\lower .1ex\hbox{\rlap{\raise .6ex\hbox{\hskip .3ex
	{\ifmmode{\scriptscriptstyle <}\else
		{$\scriptscriptstyle <$}\fi}}}
	\kern -.4ex{\ifmmode{\scriptscriptstyle \sim}\else
		{$\scriptscriptstyle\sim$}\fi}}}

\newcommand{\cutt}[1]{\textcolor{blue}{}}

\newcommand{\Ms}{{\ensuremath{M_{\odot} }}}

\newcommand{\C}{{\ensuremath{^{12}\mathrm{C}}}}

\newcommand{\N}{{\ensuremath{^{14} \mathrm{N}}} }

\shorttitle{$z >$ 6 SMBH NIR Emission}
\shortauthors{Latif et al.}

\begin{document}

\title{{\em Euclid} and {\em Roman} with {\em JWST} Could Reveal Supermassive Black Holes at up to $z \sim$ 15}

\author{Muhammad A. Latif}

\affiliation{Physics Department, College of Science, United Arab Emirates University, PO Box 15551, Al-Ain, UAE (latifne@gmail.com)}

\author{Daniel J. Whalen}

\affiliation{Institute of Cosmology and Gravitation, Portsmouth University, Dennis Sciama Building, Portsmouth PO1 3FX}

\begin{abstract}

Although supermassive black holes (SMBHs) are found at the centers of most galaxies today, over 300 have now been discovered at $z >$ 6, including UHZ1 at $z = 10.1$ and GHZ9 at $z =$ 10.4.  They are thought to form when 10$^4$ - 10$^5$ \Ms\ primordial stars die as direct-collapse black holes (DCBHs) at $z \sim$ 20 - 25.  While studies have shown that DCBHs should be visible at birth at $z \gtrsim$ 20 in the near infrared (NIR) to the {\em James Webb Space Telescope} ({\em JWST}), none have considered SMBH detections at later stages growth down to $z \sim$ 6 - 7. Here, we present continuum NIR luminosities for a BH like ULAS J1120+0641, a $1.35 \times 10^9$ \Ms\ quasar at $z =$ 7.1, from a cosmological simulation for {\em Euclid}, {\em Roman Space Telescope} ({\em RST}) and {\em JWST} bands from $z =$ 6 - 15.  We find that {\em Euclid} and {\em RST} could detect such BHs, including others like UHZ1 and GHZ9, at much earlier stages of evolution, out to $z \sim$ 14 - 15, and that their redshifts could be confirmed spectroscopically with {\em JWST}.  Synergies between these three telescopes could thus reveal the numbers of SMBHs at much higher redshifts and discriminate between their evolution pathways because {\em Euclid} and {\em RST} can capture large numbers of them in wide-field surveys for further study by {\em JWST}.  

\end{abstract}



\section{Introduction}

SMBHs at $z >$ 6 were first detected in the \textit{Sloan Digital Sky Survey} over twenty years ago \citep{fan03}.  Since then, their numbers have risen to over 300, with 13 at $z >$ 7 \citep{mort11,ban18,yang20,wang21}.  A new class of active galactic nuclei (AGNs) has now been found at even higher redshifts, such as a $4 \times 10^7$ \Ms\ BH in UHZ1 at $z = 10.1$ \citep{cet23,akos23,gould23} and an $8\times10^7$ \Ms\ BH in GHZ9 at $z =$ 10.4 \citep{Atek23,Cast23,Kov24}.  In principle, low-mass Population III (Pop III) stars could be the seeds of these BHs if they accrete continuously at the Eddington limit or grow by super-Eddington accretion \citep{vsd15,inay16,lupi24}.  However, this scenario is problematic because Pop III BHs are born in low densities that preclude rapid initial growth \citep[e.g.,][]{ket04,wan04,latif22a} and they can be ejected from their host halos during collapse \citep{wf12}. When they do accrete, they drive gas out of their halos because of their shallow gravitational potential wells, so they are restricted to fairly low duty cycles \citep{srd18}.  

It is generally thought that the $z >$ 6 quasars grew from DCBHs that form and then rapidly grow in the low-shear environments of rare, massive halos fed by strong accretion flows \citep{ten18,smidt18,lup21,vgf21}.  The highly supersonic turbulence in these unusual halos produces supermassive stars \citep[SMSs;][]{hos13,tyr17,hle18b,tyr21a,herr23a,nan24d} without any need for exotic environments \citep[or even atomic cooling, as had been supposed for nearly 20 years;][]{latif22b}.  DCBHs grow at much higher rates than Pop III BHs because Bondi-Hoyle accretion rates scale as $M_{\mathrm{BH}}^2$, and they form in much higher densities \citep[e.g.,][]{pat23a} in more massive halos that retain gas even when it is photoionized by X-rays \citep{titans,latif21a}.  SMSs are also probably required to explain the large nitrogen excesses in GN-z11 \citep{bun23,sench24}, CEERS 1019 \citep{lar23}, and GS 3073 \citep{ji24}, high-redshift galaxies at $z \sim 5.5 - 10.5$ \citep{nag23b,nan24c,nan25a}.

SMSs can be detected in the NIR at $z \sim$ 8 - 14 \citep{sur18a,sur19a,vik22a,nag23a} and DCBHs can be found at $z \gtrsim$ 20 by {\em JWST} \citep{nat17,bar18,wet20b} and at $z \sim$ 8 - 10 by the Square Kilometer Array (SKA) and next-generation Very Large Array \citep[ngVLA;][]{wet20a,wet21a}.  Quasars like ULAS J1120+0641, a $1.35 \times 10^9$ \Ms\ BH at $z =$ 7.1 that is typical of $z \sim$ 6 - 7 SMBHs, can be detected at much earlier stages of evolution ($z \sim$ 14 - 15) at 0.1 - 10 GHz, but only in blind surveys that may not yield many objects because of these telescopes' small fields of view \citep{wet23,latif24b,latif24a,latif25a,latif25b}.  

However, {\em Euclid} and {\em RST} could in principle photometrically identify much larger numbers of SMBHs at $z >$ 6 - 7 because of their large survey areas.  Once found, their properties and redshifts could be determined spectroscopically by {\em JWST}.  Observations of SMBHs at earlier stages of evolution at $z >$ 7 are clearly needed to determine their numbers and probe their formation pathways.  We have calculated NIR luminosities for a quasar like J1120 at earlier stages of evolution, $z =$ 6 - 15, to determine at what redshifts it could be found by {\em Euclid}, {\em RST}, and {\em JWST}.  In Section 2 we describe our NIR AB magnitude estimates. We  discuss AB mag limits for the quasar in redshift for all three telescopes in Section 3.    

\begin{figure}
\plotone{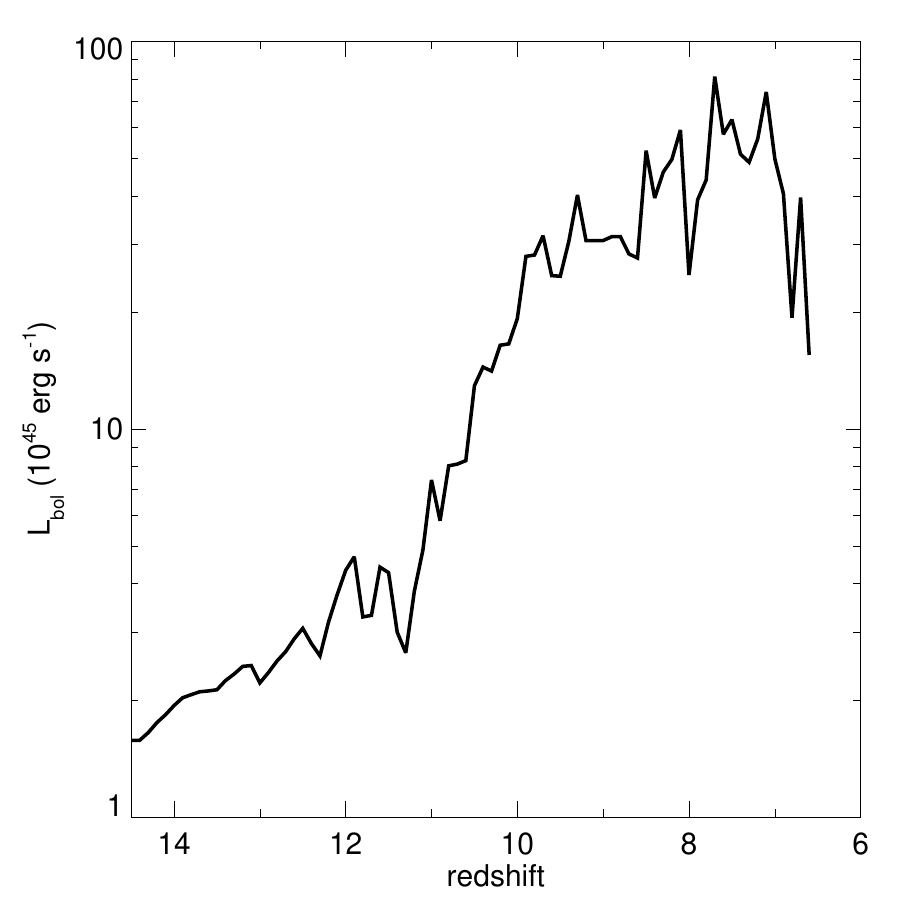}
\caption{Bolometric luminosities, $L_{\mathrm{bol}}$, for the quasar as a function of redshift from \citet{smidt18}.}
\label{fig:Lbol}
\end{figure}

\section{Numerical Method}

We first calculate rest-frame spectra for our fiducial quasar by normalizing theoretical spectra for BH accretion disks from \citet{yue13} to bolometric luminosities for the quasar in the cosmological simulation by \citet{smidt18}.  This simulation used radiation hydrodynamics to model X-ray feedback from the BH and ionizing UV from stars in the host galaxy along with local supernova feedback and chemical enrichment due to massive stars. It produced a quasar with the same mass and star formation rate as in J1120 at $z =$ 7.1. We utilize the data from this simulation to compute spectra and then cosmologically dim and redshift these spectra and convolve them with {\em Euclid}, {\em RST}, and {\em JWST} filter functions to obtain AB magnitudes.  Bolometric luminosities, $L_{\mathrm{bol}}$, for the quasar are shown for $z =$ 6 - 14.5 in Figure~\ref{fig:Lbol}.  Masses and accretion rates for the BH are shown in Figure~2 of \citet{smidt18}.  They vary from 10$^6$-10$^9$ \Ms\ and 0.01-20 \Ms\ yr$^{-1}$ ($\sim$ 0.2 - 0.8 $\dot{M}_{\mathrm{Edd}}$) from $z =$ 15 - 7.  At these rates the disk is expected to be thin, Compton-thick and radiatively efficient, so the spectral luminosity can be modeled as the sum of three components \citep{yue13},
\begin{equation}
L_{\nu} = L_{\nu}^{\mathrm{MBB}} + L_{\nu}^{\mathrm{PL}} + L_{\nu}^{\mathrm{refl}},
\end{equation}
a multicolor blackbody part due to the range of temperatures across the disk, a power-law component from the surrounding hot corona, and a reflection component, respectively.  We omit the contribution due to reflection because it is at most 10\% of the entire spectrum and is at energies well above those redshifted into the NIR today.

The temperature of the BH accretion disk is highest at its center, where
\begin{equation}
T_{\mathrm{max}} = \left(\frac{M_{\mathrm{BH}}}{M_{\odot}}\right)^{-0.25} \mathrm{keV},
\label{eq:Tmax}
\end{equation}
and decreases at larger radii \citep{mak00,shf05}.  As the SMBH grows from 10$^6$ - 10$^9$ \Ms\ temperatures at the center of the disk fall from $\sim$ 600,000 K to 60,000 K.  Equation~\ref{eq:Tmax} is valid if the central engine is a Schwarzschild BH, the innermost radius is about five times the Schwarzschild radius and the accretion is at or near the Eddington limit. The spectral luminosity is then
\begin{equation}
L_{\nu}^{\mathrm{MBB}} = L_{\mathrm{MBB}} \int_0^{T_{\mathrm{max}}} B_{\nu}(T) \left(\frac{T}{T_{\mathrm{max}}}\right)^{-11/3} \frac{dT}{T_{\mathrm{max}}},
\end{equation}
where $B_{\nu}(T)$ is the emission spectrum of a blackbody with temperature $T$ and $L_{\mathrm{MBB}}$ is a normalization factor.
 
As in \citet{yue13}, we parametrize the emission spectrum of the hot corona as a power law with an exponential cutoff:
\begin{equation}
L_{\nu}^{\mathrm{PL}} = L_{\mathrm{PL}} \, \nu^{\alpha_{\mathrm{s}}} \, \mathrm{exp}(-h\nu/E_{\mathrm{cut}}),
\end{equation}
where we set $\alpha_{\mathrm{s}} =$ 1, $E_{\mathrm{cut}} =$ 300 keV \citep{sos04}, and $L_{\mathrm{PL}}$ is a normalization factor.  As in \citet{shf05} and \citet{yue13}, we truncate the power law below the peak of the disk component, at $\sim$ $3 \,T_{\mathrm{max}}$.

We estimate normalization factors $L_{\mathrm{MBB}}$ and $L_{\mathrm{PL}}$ by setting 
\begin{equation}
\int (L_{\nu}^{\mathrm{MBB}}+ L_{\nu}^{\mathrm{PL}}) d\nu = L_{\mathrm{bol}}.  
\end{equation}
Following \cite{yue13}, this then leads to
\begin{equation}
L_{\mathrm{MBB}}= 0.5 \, L_{\mathrm{bol}}/\int L_{\nu}^{\mathrm{MBB}} d\nu
\end{equation}
and 
\begin{equation}
L_{\mathrm{PL}}= 0.5 \, L_{\mathrm{bol}}/\int L_{\nu}^{\mathrm{PL}} d\nu.
\end{equation}
\begin{figure}
\plotone{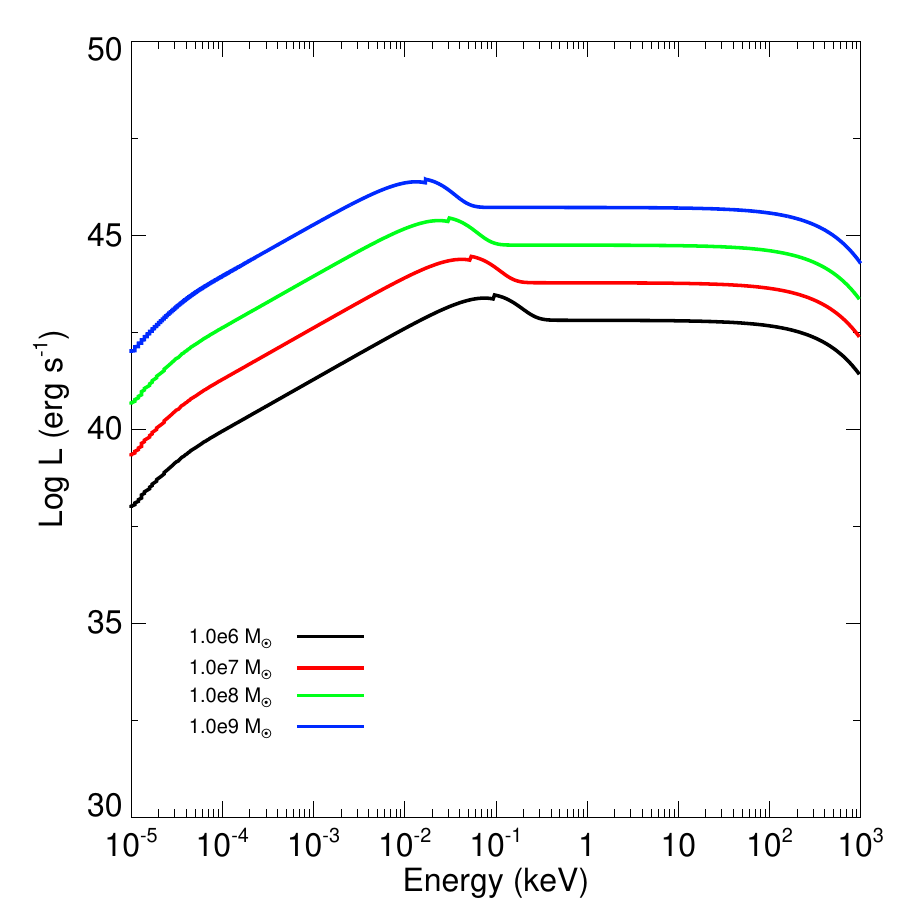}
\caption{Rest frame quasar spectra at $10^6$, $10^7$, $10^8$ and $10^9$ \Ms, the green, blue, purple and red lines, respectively.}
\label{fig:SED}
\end{figure}
We show spectra at $10^{-5}$ -$10^3$ keV over four decades in BH mass, 10$^6$ - 10$^9$ \Ms, in Figure~\ref{fig:SED}.  As the mass increases, the spectral peak shifts to lower energies, resulting in a softening of the quasar spectrum.  This is consistent with the decrease in $T_{\rm max}$ by about an order of magnitude over this range.  However, the slope of the spectrum is nearly constant over the interval in energy that is redshifted into the NIR bands we consider. 

\section{Results}

\begin{figure} 
\center
\begin{tabular}{c}
\epsfig{file=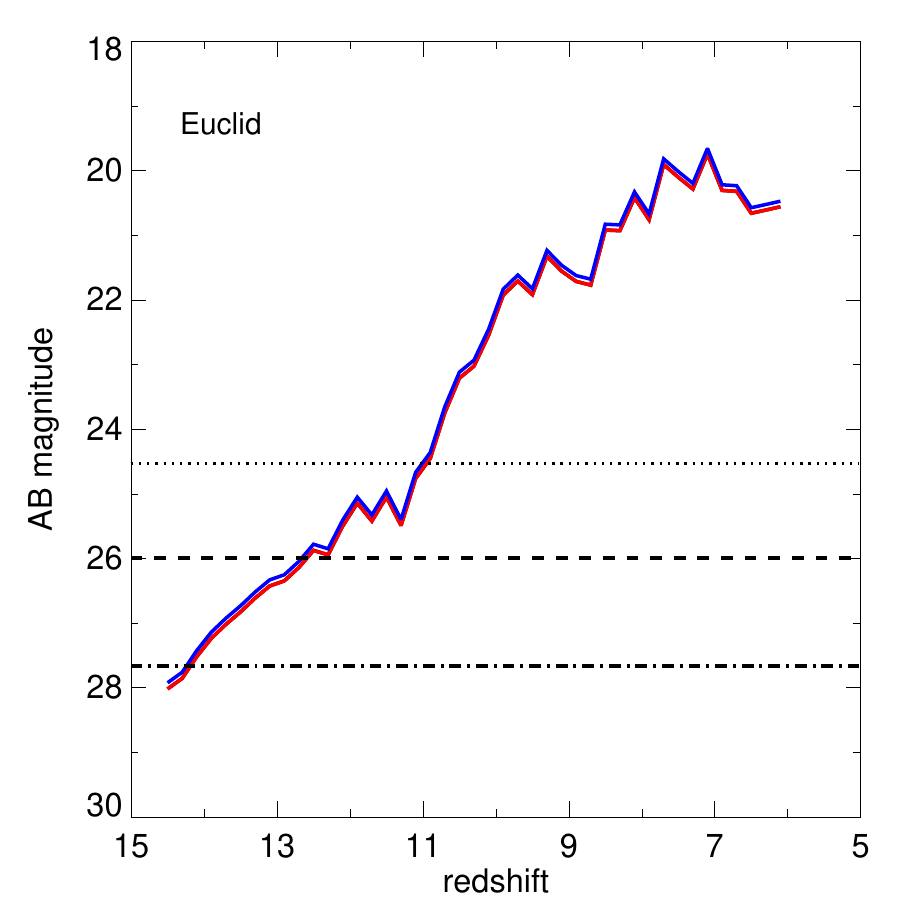,width=0.75\linewidth,clip=}  \\
\epsfig{file=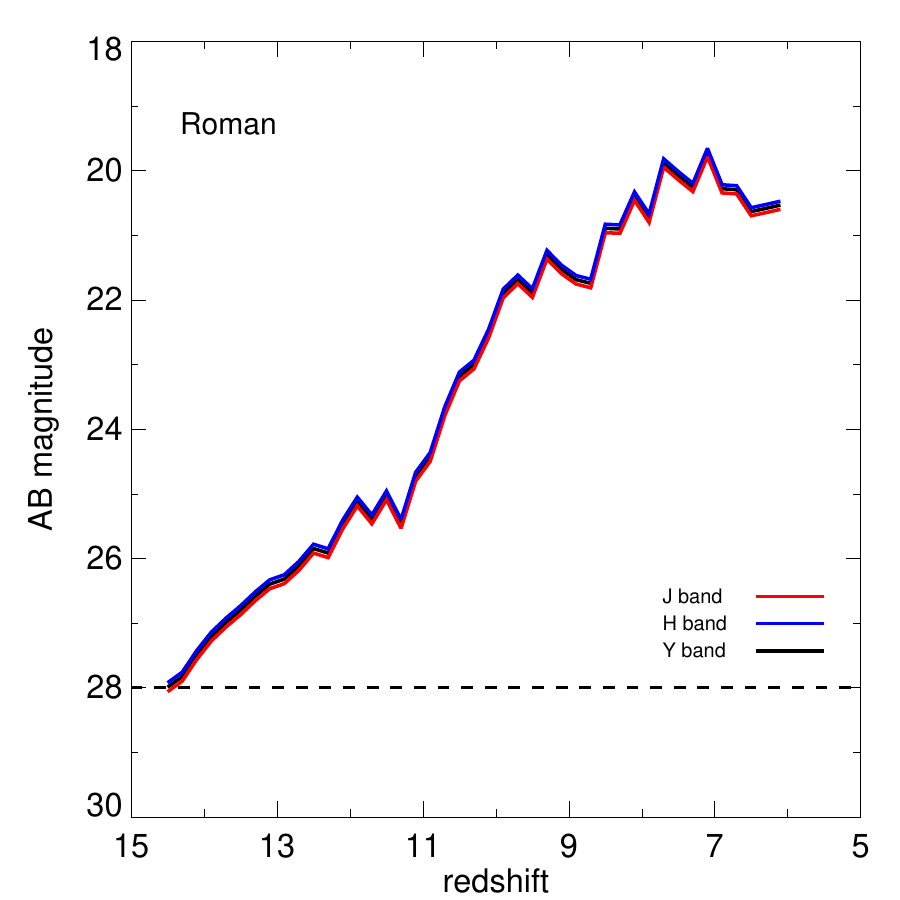,width=0.75\linewidth,clip=}  \\
\epsfig{file=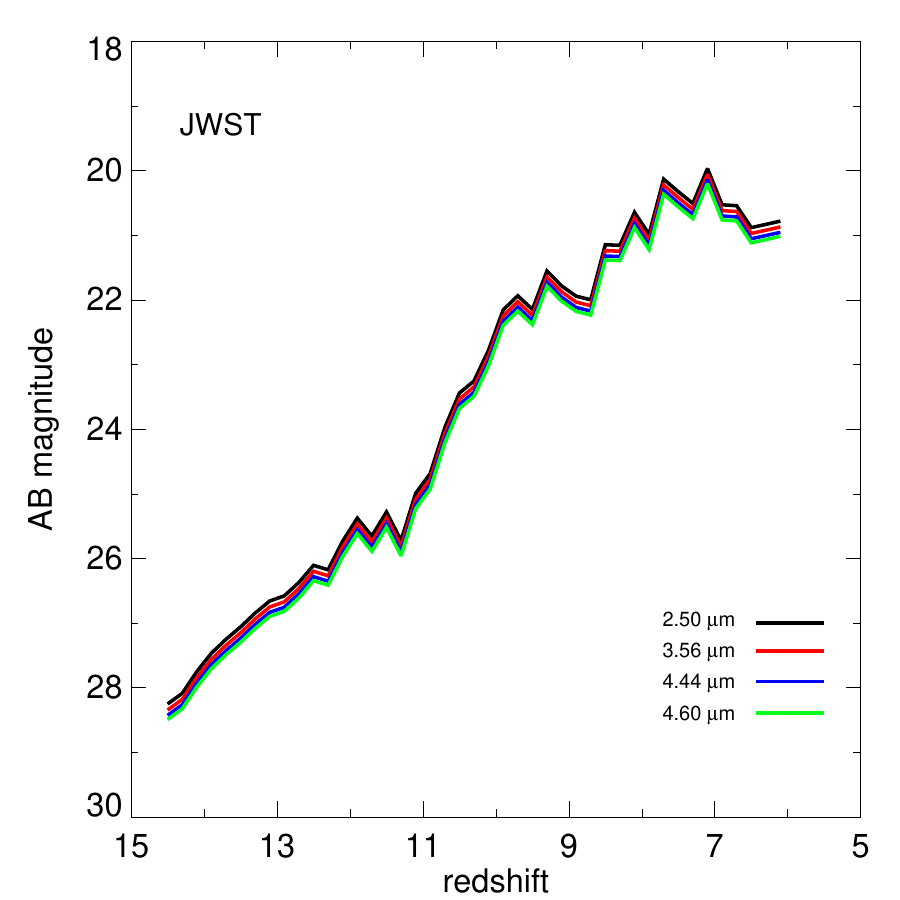,width=0.75\linewidth,clip=}  
\end{tabular}
\caption{Top:  AB magnitudes in \textit{Euclid} \textit{H} and \textit{J} bands.  The horizontal dotted, dashed and dot-dash lines are AB mag limits for the Euclid Wide, Deep and Ultra Deep Surveys, respectively.  Center:  AB magnitudes in \textit{RST} \textit{H}, \textit{J} and \textit{Y} bands.  The horizontal dashed line is the AB mag limit for the High Latitude Spectroscopic Survey.  Bottom:  AB magnitudes in the \textit{JWST} 2.50 $\mu$m, 3.65 $\mu$m, 4.44 $\mu$m and 4.60 $\mu$m NIRcam bands.  The \textit{JWST} NIRcam AB magnitude limit is 32 for these filters, below the scale of the plot.
\label{fig:abmag}}
\end{figure}

We show AB magnitudes for the quasar at $z =$ 6 - 14.5 in the \text{Euclid} \textit{H} and \textit{J} bands and the \textit{RST} \textit{H}, \textit{J} and \textit{Y} bands in the top and center panels of Figure~\ref{fig:abmag}  (although we note that \text{RST} could detect SMBHs out to $z \sim$ 18 because its F213 filter extends down to 2.30 $\mu$m, unlike \textit{Euclid'}s $H$ band filter, which cuts off at 2.0 $\mu$m and would limit detections to $z \sim$ 15.5).    We superimpose detection limits for the 15000 deg$^2$ Euclid Wide Survey \citep[WS, 24.5 AB mag;][]{euclid2}, the 50 deg$^2$ Deep Survey (DS; 26.0 AB mag), the $\sim$ 1 deg$^2$ Ultra Deep Survey (EUDS; 27.7 AB mag) on the top panel and for the 2000 deg$^2$ \textit{RST} High Lattitude Spectroscopic Survey \citep[HLSS, 28 AB mag;][]{HLSS} in the middle panel.  The \textit{Euclid} WS could photometrically detect a quasar like J1120 at redshifts up to $z \sim$ 11 while the DS and EUDS could detect them out to $z \sim$ 13 and $z \sim$ 14, respectively, when the BH only has a mass of $\sim$ 10$^6$ \Ms\ and an accretion rate of $\sim$ 0.8 $\dot{M}_{\rm Edd}$ \citep[see Figures 2 and 3 of][]{smidt18}.  These are effectively the upper limits in $z$ for detection by \textit{Euclid} because photons redshifted into the \textit{H} and \textit{Y} band filters from earlier times would be blueward of the Lyman limit in the rest frame and be resonantly scattered or absorbed by the neutral IGM.  

In contrast, \textit{RST} could detect this quasar at $z \lesssim$ 14.5 in the \textit{H}, \textit{J} and \textit{Y} filters, making it the observatory of choice for initial detection of quasars at earlier times because of its high sensitivities and large survey areas.  We show AB magnitudes in the 2.50 $\mu$m, 3.65 $\mu$m, 4.44 $\mu$m and 4.60 $\mu$m \textit{JWST} NIRcam filters in the bottom panel of Figure~\ref{fig:abmag}.  They are well above NIRcam detection limits ($\sim$ AB mag 32, not visible in the plot), or even NIRSpec limits, $\sim$ 29 at $z \lesssim$ 15.  \textit{JWST} could thus spectroscopically confirm redshifts of BHs with masses as low as $\sim$ 10$^6$ \Ms\ if they accrete at or about the Eddington limit at $z \sim$ 15.  \citet{Cast23} reported NIR AB magnitudes of 26.2 for UHZ1 and 27.1 for GHZ9 in the JWST F200 filters so the Euclid Deep and Ultradeep Surveys could detect UHZ1 and the RST HLSS could detect either object at their respective redshifts.  Since the inferred masses of these two BHs are $\sim$ 10$^7$ \Ms\ and $7 \times 10^7$ \Ms, respectively, our NIR magnitudes indicate that they also could have been detected at earlier stages of growth at masses $\gtrsim$ 10$^6$ \Ms, likely out to $z \sim 15$.

\begin{figure}
\plotone{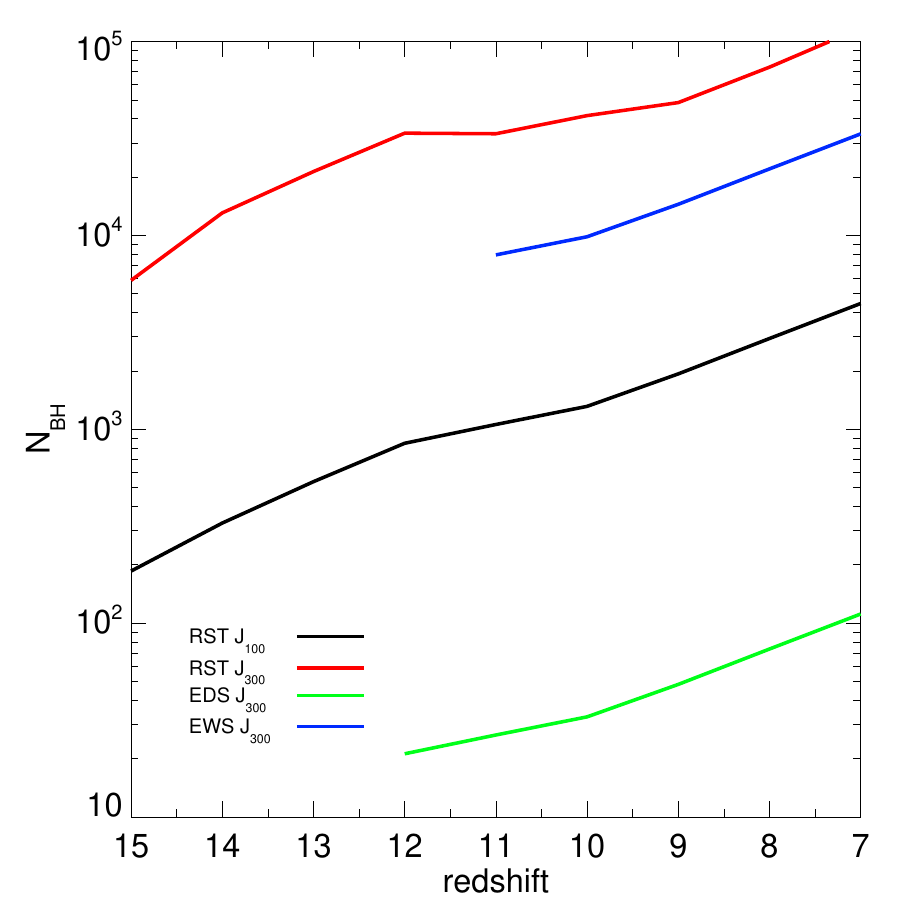}
\caption{Expected numbers of BH detections in the ED, EUD and HLS surveys based on DCBH number densities from simulations with UV backgrounds of 100 $J_{21}$ and 300 $J_{21}$, where $J_{21}$ = 10$^{-21}$ erg s$^{-1}$ cm$^{-2}$ Hz$^{-1}$ sr$^{-1}$.  From {\em JWST} observations of AGNs that suggest number densities of 10$^{-5}$ Mpc$^{-3}$ at $z =$ 10 we expect up to 2000 and 60,000 BHs in the EDS and HLSS at this redshift, respectively.}
\label{fig:nbh}
\end{figure}

Predictions of early quasar counts in survey fields depend on their number densities at high redshift, which are uncertain, and survey depth.  On the low end, large-scale cosmological simulations dedicated to high-$z$ BH growth indicate that there are about a dozen gas reservoirs per Gpc$^{3}$ that are capable of hosting 10$^9$ \Ms\ quasars by $z \sim$ 7 \citep[e.g.,][]{dm12,dm17}.  On the high end, predictions of DCBH number densities vary from 10$^{-5}$ - 10$^{-8}$ Mpc$^{-3}$ at $z =$ 10 - 15, depending on the strength of UV backgrounds \citep{hab16,ob25}, so clearly not all of them are destined to become quasars because their growth is quenched at some point.  Most will instead become less luminous AGNs such as the 'little red dots' \citep{gould23,koc25}.  

We show the expected numbers of BH detections in the EW, ED, EUD and HLS surveys in Figure~\ref{fig:nbh} based on predicted numbers of DCBHs from simulations as a function of redshift and for the {\em JWST} AGN luminosity function \citep{kok24,koc25,inay25} at $z =$ 10, which is on par with those of DCBHs.  From simulations we find that {\em RST} could detect $\sim$ 200 - 6000 BHs at $z \sim$ 15 while the {\em Euclid} DS is limited to $\sim$ 20 at $z =$ 12.5.  No detections are expected for the EUDS because its small field (1 deg$^2$) is too small to enclose any objects, in spite of its high sensitivity.  For \textit{JWST} luminosity functions of 10$^-5$ Mpc$^-3$ for high-$z$ AGNs (Figure 4 of \citealt{Kov24} and Figure 2a of \citealt{inay25}), the numbers for EDS and HLSS  are predicted to be $\sim$ 2000 and 60,000, respectively.  

For the range of DCBH number densities predicted by simulations, from $\sim$ 1000 - 30,000 BHs like UHZ1 or GHZ9 could be found at $z =$ 10.5 down to limiting magnitudes of 28 in HLSS and $\sim$ 30 BHs like UHZ1 could be nominally be detected in the EDS at the same redshift at survey depths of 26.  Extrapolations of quasar luminosity functions \citep[QLFs;][]{barn19} predict that Euclid should detect 6 - 8 quasars at $z > 8$ with magnitudes greater than 23 and and that RST should find 180 quasars at $6.5 < z < 9$ and more than 20 at $z>7.5$ over 2000 deg$^2$ \citep{tee23}.  These numbers are lower than ours because we consider all BHs, not just the most massive ones that become quasars. 

We exclude contributions due to stars in the host galaxy in our SMBH spectra but can guage their effect on AB magnitudes by comparison to \citet{nat24}, who include them in their spectral templates for UHZ1.  The stars in their synthetic spectra come from analytical fits to an evolving stellar population in a cosmological simulation \citep{agarw14} but they assume the same spectra for the BH as ours so we can determine the effect of the absence of the host galaxy in our AB magnitudes.  They obtain a H band AB magnitude $\sim$ 26.5, which is lower than our $\sim$ 23 at the same redshift but their BH was a factor of 7 lower in mass.  When this fact is taken into account the disparity due to stars in the host galaxy is $\sim$ 1 - 1.5 AB mag, which is within the uncertainty of the observed luminosity and does not change our core result.  Dust, which has been observed in high-redshift quasars \citep[e.g.,][]{maio04}, can also reprocess flux from the BH to longer wavelengths but is subject to radiation pressure that rapidly drive it from the galaxy \citep[e.g.,][]{bieri17,costa18a}.  Nonetheless, the influence of stars and dust in host galaxies of primordial quasars should be quantified in future work.

\section{Discussion and Conclusion}

\textit{RST} will have unparalleled sensitivity over large survey areas so it will be the best bet for initial identification of high-$z$ quasars.  It also has seven filters as opposed to \textit{Euclid'}s four for better initial redshift cuts for dropouts.  Although \textit{RST} will perform slitless spectroscopy at 0.8 - 1.9 $\mu$m, the one hour exposure required to reach AB mag 28 in the \textit{H} band will only produce 10$\sigma$ continuum sensitivity limits of AB mag 21 and 23 for point sources in the prism and grism, respectively, insufficient for determination of high redshifts.  Consequently, both \textit{RST} and \textit{Euclid} will have to rely on efficient photometric techniques such as dropouts or fits of spectral energy distribution templates to the observed photometry with Bayesian methods or neural networks to determine redshifts at high $z$.  The \citet{barn19} compared Bayesian and minimal-$\chi^2$/SED fitting methods and found that Bayesian methods, which can reach to at least 0.5 magnitude fainter, work better than $\chi^2$ fitting method at lower redshifts ($ 7 < z < 8$) but at higher redshifts ($z > 8$) they yield similar results (see their Table 3).

However, the accuracy of these techniques can fall off at high redshifts, so \textit{JWST} or ground-based extremely large telescopes (ELTs) are needed to spectroscopically confirm redshifts for high-$z$ SMBH candidates found by \textit{RST} and \textit{Euclid}.  \textit{JWST} is the best telescope for this purpose, given the (sometimes severe) systematics due to atmospheric transmission associated with the ELTs.  SKA and ngVLA could also play an important role by confirming the presence of a BH in these candidates at $z \lesssim$ 14, especially at 1 - 10 GHz where flux from the BH dominates synchrotron emission by supernova remnants in the host galaxy \citep[see Figure 2 in][]{latif24a}.  

\textit{Euclid}, \textit{RST} and \textit{JWST} will for the first time directly probe the the era of SMBH evolution from $z \sim$ 5 - 15.  Such observations will be key to understanding how populations of early quasars broke off from less massive populations such as AGNs and IMBHs at early redshifts, given the great disparity of their numbers at $z =$ 6 - 7, $\sim$ 1 Gpc$^{-3}$ versus a few 10$^{-5}$ Mpc$^{-3}$.  Followup observations on these BHs and their host galaxies will yield insights into how feedback from the BH and ionizing UV and supernovae from stars regulated SMBH growth.  Exciting new synergies between \textit{Euclid}, \textit{RST} and \textit{JWST} will inaugurate the era of $z \lesssim$ 15 BH astronomy in the coming decade.

\acknowledgments

We thank the anonymous referee for their comments, which improved this letter.  MAL acknowledges the UAEU for funding via UPAR grants No. 31S390 and 12S111.  

\bibliographystyle{apj}
\bibliography{refs}

\end{document}